\documentclass[a4paper,11pt]{article}
\usepackage{jheppub}
\usepackage{amsmath,amssymb,graphicx}
\usepackage{lineno}
\usepackage[normalem]{ulem}
\usepackage{appendix}
\newcommand{\phast}{\textbf{}{PhAST}}
\newcommand{\npevst}{n_{\mathrm{PE}}(t)}

\title{\boldmath \phast: Simultaneous reconstruction of photoelectron count and time profiles from PMT waveforms via machine learning
}

\author[a]{Yiming Xu}
\author[b]{Youwen Fan}
\author[c]{Siyu Chen}
\author[b]{Hongyue Duyang}
\author[b]{Teng Li}
\author[d]{Yaoguang Wang}
\affiliation[a]{Taishan College, Shandong University, Jinan, Shandong, China}
\affiliation[b]{Institute of Frontier and Interdisciplinary Science, Shandong University,\\Qingdao, Shandong, 266237, China}
\affiliation[c]{School of Computer Engineering, Jimei University,\\Xiamen, Fujian, China}
\affiliation[d]{School of Physics, Shandong University, Jinan, Shandong, China}

\emailAdd{xuyiming@mail.sdu.edu.cn}
\emailAdd{fanyw@mail.sdu.edu.cn}
\emailAdd{chensy@ieee.org}
\emailAdd{duyang@sdu.edu.cn}
\emailAdd{tengli@sdu.edu.cn}
\emailAdd{wangyaoguang@sdu.edu.cn}

\abstract{
Photomultiplier tubes (PMTs) are widely used in particle and nuclear physics experiments.
The reconstruction of PMT waveforms is a fundamental task in these experiments, 
where accurate extraction of photoelectron (PE) multiplicities and time from the waveform is required for downstream event reconstruction and analysis. 
In realistic detector environments, PMT waveform reconstruction is complicated by electronic effects such as pileup, charge fluctuations, noise etc., which make precise recovery of physical observables challenging.
To address these challenges, we present \phast{}, a machine-learning-based method that reconstructs PE count and time profile simultaneously.  
The model consists of a shared wave-transformer encoder followed by two dedicated branches: a counting branch for the total PE number prediction,  
and a time branch employing a count-conditioned query decoder with dynamic query activation.
To study the reconstruction performance under controlled conditions, we construct several toy Monte Carlo PMT waveform datasets, including both uniform and mixed fast-slow double-temporal-components configurations.
The proposed method demonstrates stable and accurate reconstruction performance across various waveform conditions, achieving high consistency in both PE counting and time reconstruction. 
These results indicate that architectures combining convolutional feature extraction with query-based transformer decoders provide an effective approach for complex PMT waveform reconstruction tasks.
}

\begin{document}
\maketitle
\flushbottom

\section{Introduction}
\label{sec:intro}
Photomultiplier tubes (PMTs) are widely used in current and next-generation neutrino and dark matter experiments for the detection of scintillation and Cherenkov light, providing essential measurements of photon arrival times and intensities. 
%A wide range of contemporary and next-generation experiments rely on PMT-based readout systems. 
%In neutrino physics, examples include the liquid scintillator detectors such as KamLand~\cite{}, Borexino~\cite{}, Daya Bay~\cite{}, and JUNO~\cite{}, water Cherenkov detectors including Super-Kamiokande~\cite{} and Hyper-Kamiokande~\cite{}, and the liquid argon time projection chambers that are to be used by DUNE~\cite{}. 
In neutrino experiments, water Cherenkov detectors such as Super-Kamiokande~\cite{Fukuda2003SuperK}, Hyper-Kamiokande~\cite{Abe2021HyperK} instrument thousands to tens of thousands of PMTs to detect Cherenkov light from neutrino interactions. 
Jiangmen Underground Neutrino Observatory (JUNO) with approximately 17,000 20-inch PMTs and 25,000 3-inch PMTs represents a flagship example of liquid scintillator detectors that play an important role in neutrino physics~\cite{an2016juno,Abusleme2022JUNOelectronics}. 
%LArTPC...? ICECUBE, KM3NeT...
In dark matter direct detection, liquid xenon time projection chambers like XENONnT~\cite{Aprile2024XENONnT}, LZ~\cite{Akerib2020LZ}, and PandaX-4T~\cite{Zhang2019PandaX4T} use PMT arrays to record both scintillation light and electroluminescence signals for 3D event localization and background discrimination. 
%In gamma-ray astronomy, imaging atmospheric Cherenkov telescopes (IACTs) such as VERITAS, H.E.S.S., and the upcoming Cherenkov Telescope Array (CTA) employ fast PMTs to capture nanosecond-scale Cherenkov flashes from extensive air showers. 
%Beyond these, PMTs are also fundamental in time-of-flight positron emission tomography (TOF-PET), liquid scintillator detectors in reactor neutrino experiments, and rare decay searches such as Mu2e and COMET. 

Waveform reconstruction is one of the central tasks in many of these experiments employing PMTs. Its goal is typically to extract the number of photoelectrons (PEs) and their temporal structure that serve as inputs
%Accurate waveform reconstruction is critical 
to the downstream tasks including the determination of interaction vertices and event energy deposition, particle identification, and background rejection, which are fundamental for various physics analyses.
Achieving robust and accurate waveform reconstruction is therefore an important prerequisite for overall detector performance.

Over the past decades, different methods have been developed for PMT waveform reconstruction, such as template fitting, deconvolution and maximum likelihood estimation~\cite{Abusleme2022JUNOelectronics,Yu2021Deconvolution,Adamson2002MLE}.
However, under high pile-up conditions where multiple PEs are produced within a short time window, their individual pulses superimpose, making it difficult for traditional reconstruction approaches to reconstruct accurate PE number and time profile. 
%multiple PE signals but also PMT responses such as charge traditional reconstruction approaches difficult to both accurate PE number and precise time. 
%such as template matching or deconvolution unstable and susceptible to bias.

To address the reconstruction of complex PMT waveforms under high pile-up conditions, a variety of methods have been proposed, including the Fast Stochastic Matching Pursuit methods ~\cite{Xu2024FSMP} and machine-learning-based approaches~\cite{Jiang2025MLPhotonCounting,Zhang2025DiffusionPMT}. 
%However, the exploration of end-to-end reconstruction methods remains relatively insufficient. 
In this work, we formulate the \textbf{Ph}oto \textbf{A}rrival \textbf{S}equence \textbf{T}ransformer (\phast{}) model, which deals with PMT waveform reconstruction as an end-to-end structured prediction problem. 
Given an input waveform, the model reconstructs both the PE count and time profile ($\npevst$) simultaneously with precision. 
Architecturally, \phast{} is designed with transformer-based sequence encoding~\cite{vaswani2017attention} together with query-based set prediction ideas inspired by DETR-like and mask-based architectures~\cite{carion2020end,cheng2022masked}, while adapting the prediction heads, matching strategy, and evaluation metrics to one-dimensional waveform reconstruction. Rather than treating the task purely as dense regression, \phast{} explicitly models photon-arrival hypotheses and aligns predictions with ground truth through Hungarian matching~\cite{kuhn1955hungarian}.

%This paper focuses on three parts of the problem. 
This paper is organized as follows. In section~\ref{sec:data}, we construct a waveform simulation and preprocessing pipeline that preserves the temporal features relevant for reconstruction while keeping the data model explicit. Section~\ref{sec:method} designs \phast{} as a two-branch count-and-time architecture in which waveform-level total number of PE estimation regulates a count-conditioned query decoder. After the training dataset preparation and optimization described in section~\ref{sec:training},  the model performance is evaluated on a family of fast/slow double-temporal-components-mixture datasets (FastSlow) and a uniform dataset (Uniform) in order to study how the method behaves under different circumstances in section~\ref{sec:eval}, and is further discussed in section~\ref{sec:discussion}. 
Finally, section~\ref{sec:conclusion} concludes the study.

\section{Dataset}
\label{sec:data}
\subsection{Waveform simulation}

\begin{figure}[t]
\centering
\includegraphics[width=0.4\textwidth]{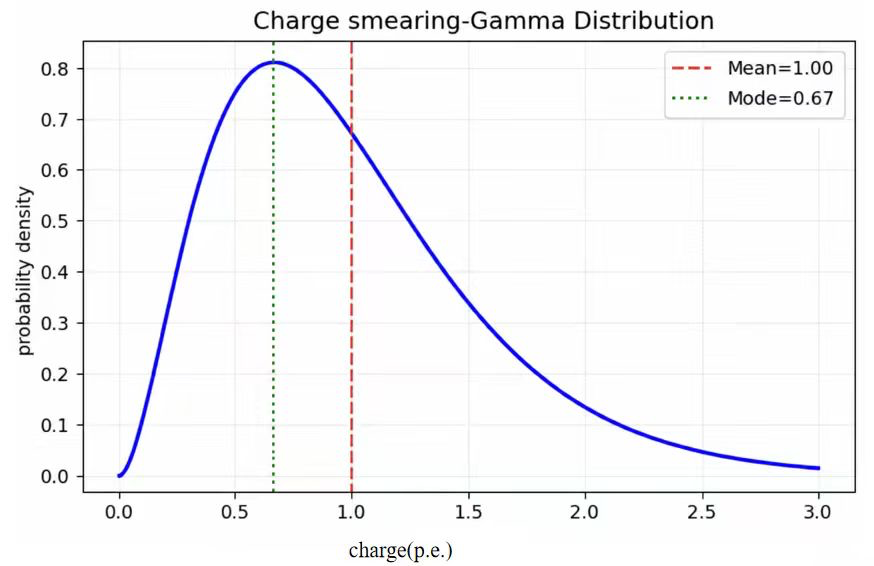}
\hfill
\includegraphics[width=0.45\textwidth]{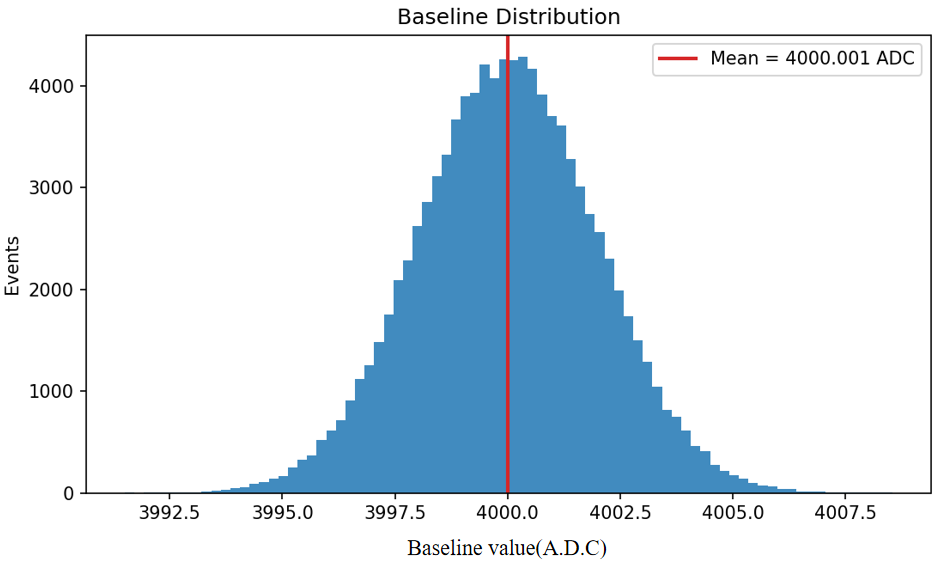}
\hfill
\includegraphics[width=0.55\textwidth]{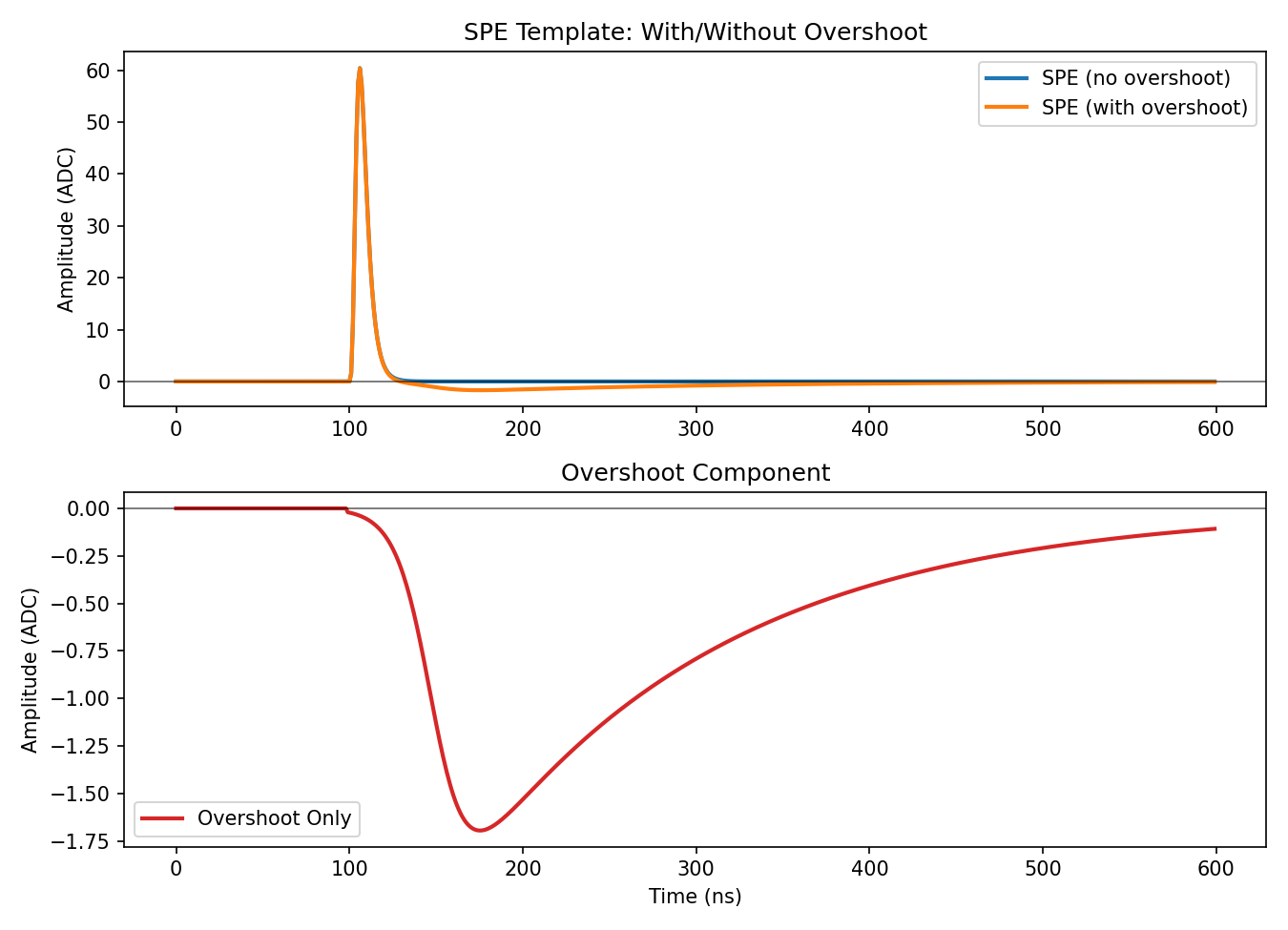}
\caption{Illustrative ingredients of the toy waveform simulation: the SPE charge distribution (top left), the baseline distribution (top right), the waveform overshoot model (bottom).}
\label{fig:toy_components}
\end{figure}

\begin{figure}[t]
\centering 
\includegraphics[width=0.45\textwidth]{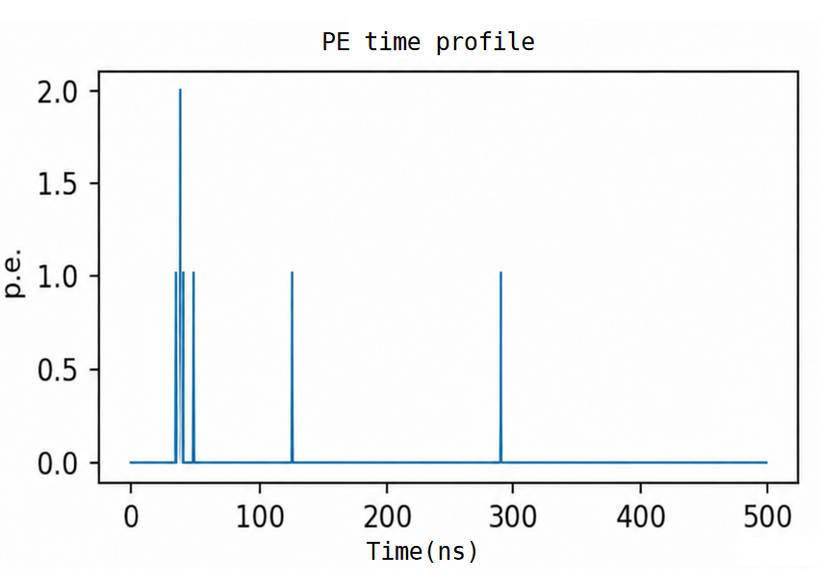} 
\vspace{1em} 
\hfill

% 第二行
\includegraphics[width=0.45\textwidth]{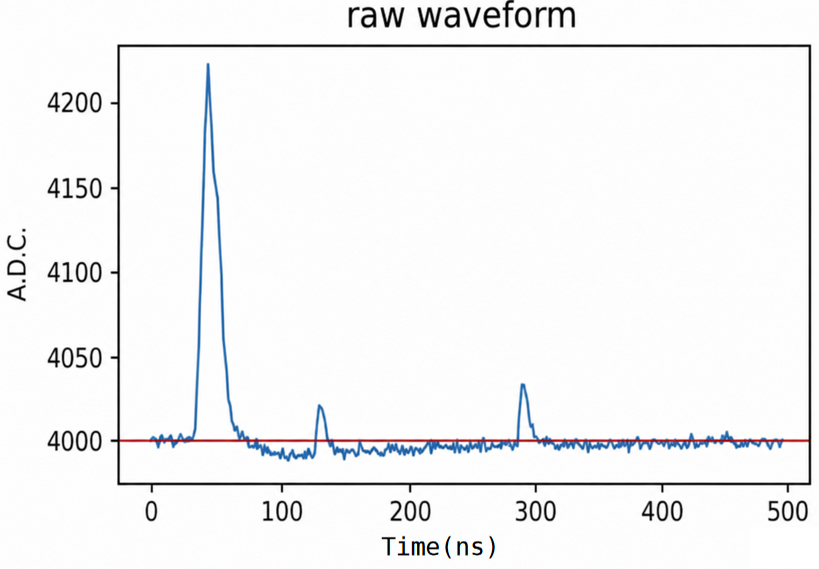}
\hfill 
\raisebox{-2pt}{\includegraphics[width=0.47\textwidth]{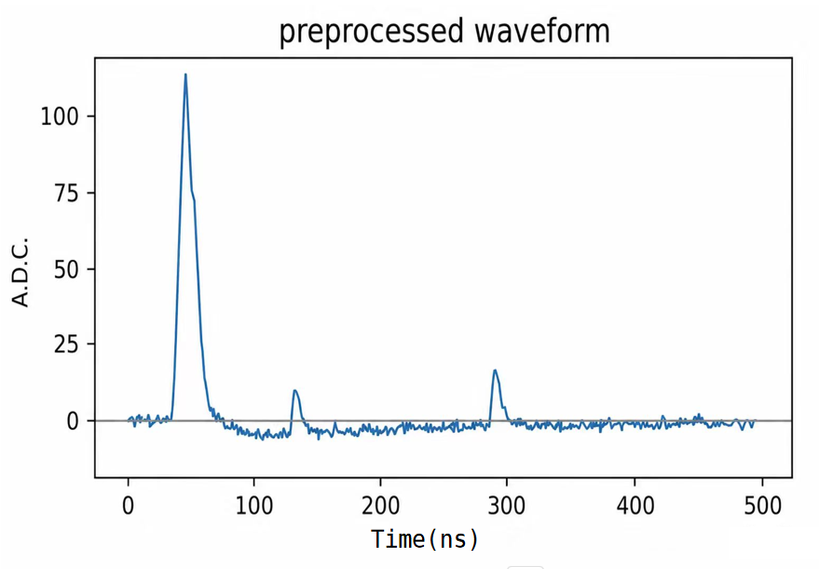}}

\caption{An example of PE time profile (top) and its corresponding raw (bottom left) and preprocessed (bottom right) waveforms used in this study. 
}
\label{fig:waveform_example}
\end{figure}
To study the waveform reconstruction problem under controlled conditions, we construct a toy Monte Carlo (MC) waveform simulation pipeline inspired by realistic PMT signal generation following the procedures described in Ref.~\cite{Jetter_2012}. The simulation includes PE generation, SPE response, charge fluctuation, transit-time spread (TTS), overshoot, baseline variation, and electronic noise.

For each event, the total PE number is randomly sampled between \(0\) and \(10\). After the PE count is determined, PE arrival times are generated in a 500\,ns time window according to predefined time distributions.

Two kinds of datasets are designed for this study with different PE time distributions. %: the \textbf{FastSlow} dataset, and the \textbf{Uniform} datasets. }
The first is the \textbf{FastSlow} datasets, in which the PE time distribution follows a two-component exponential mixture,
\begin{equation}
p(t) \propto
r e^{-t/\tau_f}
+
(1-r)e^{-t/\tau_s},
\end{equation}
where $\tau_f$ and $\tau_s$ are the time constants of the fast/slow components respectively,  with \(r=0.7\) to account for their fractions. Several combinations of fast and slow decay constants are used:
\[
(\tau_f,\tau_s)
=
(20,100),
(30,100),
(40,100),
(50,100),
(80,200)\ \mathrm{ns}.
\]
These datasets are designed to achieve different pileup conditions and time complexities.
Besides, we also construct a \textbf{Uniform} dataset in which PE times are sampled uniformly over the time window. This dataset removes intrinsic temporal structure from the generation process and serves as a simplified example.

To simulate a realistic PMT response, an TTS fluctuation of approximately \(1\)\,ns is added to the PE time. 
Each PE is then converted into an SPE waveform pulse using an asymmetric log-normal response template from Ref.~\cite{Jetter_2012}. 
Charge fluctuation is modeled using a Gamma distribution so that different PE pulses have slightly different amplitudes (Fig.~\ref{fig:toy_components}, top left).

Additional electronic effects are also included in the simulation. A delayed negative overshoot component is added after the main SPE pulse, with an overshoot ratio of approximately \(5\%\). Baseline fluctuations (Fig.~\ref{fig:toy_components}, top right) and Gaussian white noise ($\sigma = 2.1$\,ADC) are further added to mimic realistic electronic readout conditions.
The final waveform is obtained by summing all PE pulse contributions together with overshoot, baseline, and noise terms. Figure~\ref{fig:waveform_example} shows a typical example of PE-time profile (top) and its corresponding generated waveform (bottom left).

\subsection{Waveform preprocessing}

Waveforms simulated are preprocessed to make them easier for the \phast{} model to digest. 
The preprocessing pipeline is intentionally kept simple so that physically meaningful waveform structures are preserved before training.

%For each event, the waveform is cropped to a \(500\) ns readout window. 
For each waveform, the baseline and noise fluctuation are estimated from the first \(30\) ns of the waveform. 
%After baseline subtraction, 
The baseline is first subtracted from the waveform, which is then normalized using the estimated noise RMS:
\begin{equation}
\tilde{x}_t
=
\frac{x_t-\hat{b}}
{\hat{\sigma}_{\mathrm{noise}}},
\end{equation}
where \(\hat{b}\) is the baseline estimate and \(\hat{\sigma}_{\mathrm{noise}}\) is the noise normalization factor.

After preprocessing, waveform/profile pairs are visually inspected to verify that the correspondence between waveform morphology and PE time structure remains physically reasonable.
\subsection{Difficulty-aware split and sampling}

While the total PE number (\(N_{\mathrm{PE}}\)) provides a first-order measure of waveform complexity, it is not sufficient to fully characterize the difficulty of waveform reconstruction. In particular, waveforms with similar occupancy can still exhibit very different temporal structures, depending on how strongly individual PE signals overlap in time.

To better organize the dataset, we introduce a simple difficulty-aware categorization based on both occupancy and coarse temporal structure. In addition to \(N_{\mathrm{PE}}\), we consider several heuristic indicators extracted from the ground-truth PE-time profile, including the minimum separation between adjacent hits, the maximum bin occupancy, and a short-range local density measure. These quantities are combined into a single scalar score that reflects the effective level of pileup and temporal merging in each waveform.

Based on this score, waveforms are grouped into four difficulty levels:
\begin{equation}
d=0 \rightarrow \textit{easy}, \qquad
d=1 \rightarrow \textit{normal}, \qquad
d=2 \rightarrow \textit{hard}, \qquad
d=3 \rightarrow \textit{extreme}.
\end{equation}
The detailed splitting procedure is in appendix \ref{difficulty-split}.
This labeling is used purely for data organization during training and evaluation. In practice, we adopt a stratified train/validation split to ensure that all difficulty regimes are consistently represented. In addition, minibatches are sampled in a roughly balanced manner across difficulty levels, preventing training from being dominated by low-complexity waveforms, which are more frequent but less informative for learning pileup-aware reconstruction.

This difficulty-aware sampling strategy provides a more stable training distribution compared to using \(N_{\mathrm{PE}}\) alone, and helps the model better capture waveform structures arising from strong PE arrivals pileup.

\section{The \phast{} architecture}
\label{sec:method}

\subsection{Overall design}

\begin{figure}[t]
\centering
\includegraphics[width=0.98\textwidth]{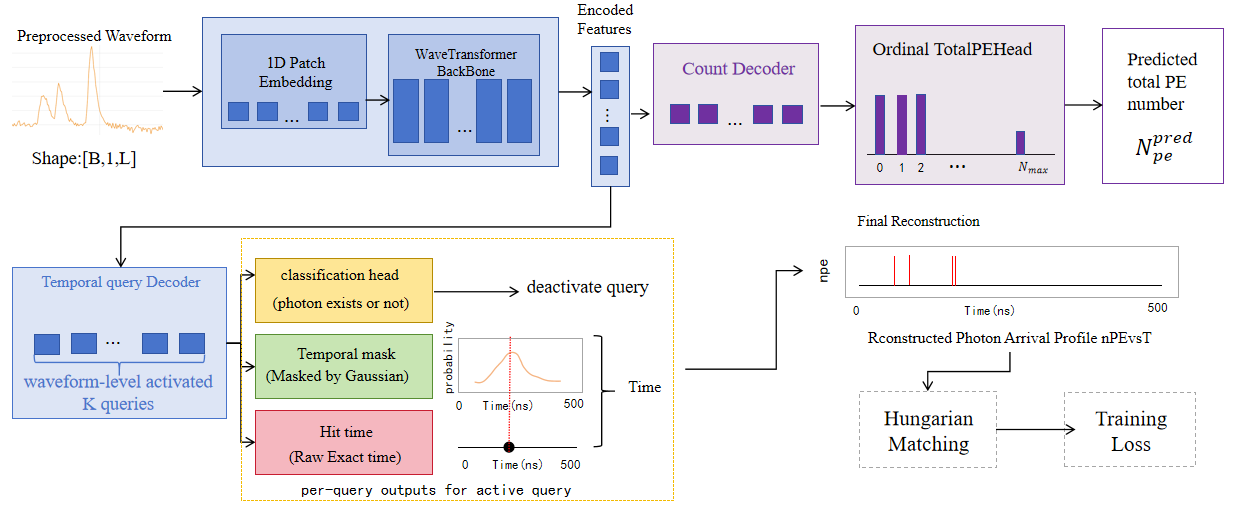}
\caption{Architecture of \phast. A preprocessed waveform is encoded by a shared transformer backbone, followed by a total PE counting branch and a count-conditioned time decoder with dynamic query activation. The accepted time queries are assembled into the reconstructed PE-time profile \(\widehat{\npevst}\).}
\label{fig:overview}
\end{figure}
\phast{} is an end-to-end waveform-to-$n_{PE}(t)$ reconstruction model. The basic idea is to split the task into two tightly coupled parts: one is to estimate how many PEs are present in the waveform, and the other is to reconstruct the time they are generated. Concretely, this corresponds to total PE number estimation via a count branch, and time-structured hit reconstruction via a count-conditioned query decoder.

The forward pass follows this idea directly. A preprocessed waveform is first encoded into a latent token sequence using a shared transformer backbone. A first decoder pass extracts features for the count branch and produces an occupancy estimate. This estimate is then fed back into the time decoder, where it determines how many queries are actually activated from a fixed-capacity query bank. The selected queries are finally aggregated into the output, reconstructing the PE-time profile \(\widehat{\npevst}\). The overall pipeline is summarized in Figure~\ref{fig:overview},  and discussed in detail in the following subsections.

\subsection{Shared waveform encoder}

The backbone of the shared waveform encoder is a one-dimensional transformer encoder. A convolutional patch embedding first maps the input waveform to a latent sequence with model dimension \(d_{\mathrm{model}}=128\). In the benchmark configuration, the backbone uses \(4\) encoder layers, \(4\) attention heads, feed-forward width \(256\), a class token, and sinusoidal positional encoding. The patch embedding is implemented with a short convolutional front-end, so the waveform is locally compressed before entering the encoder.

\subsection{Count branch: total-PE prediction}

A distinctive feature of \phast{} is that total-PE prediction is not treated as a mere auxiliary scalar regression. Instead, we perform a first decoder pass over the deep encoder sequence and send the resulting decoder hidden states to an ordinal count head named TotalPEHead. This head consists of a feature adapter, several residual one-dimensional convolutional blocks, temporal pooling, a GRU, and a final linear layer that outputs ordinal threshold logits.~\cite{Jiang2025MLPhotonCounting} Operationally, it acts as a learned occupancy estimator that summarizes the waveform before the time branch attempts dense hit reconstruction.

If the number of count classes is \(C\), the head emits \(C-1\) logits \(z_c\), corresponding to the waveforms ``\(N_{\mathrm{PE}}>c\)''. The predicted count is then
\begin{equation}
\hat{N}_{\mathrm{PE}} = \sum_{c=0}^{C-2} \mathbb{I}\!\left[\sigma(z_c)\ge 0.5\right],
\end{equation}
which is the standard inference rule for ordinal threshold targets. In the default configuration, \(C=11\), corresponding to counts from \(0\) to \(10\). This formulation stabilizes count prediction and makes it easy to reuse the predicted total PE number as a control signal for the time decoder.
And
\begin{equation}
    \mathbb{I}(A)=\begin{cases}
        1& A~\text{is true}\\
        0 & \text{others}
    \end{cases}
\end{equation}
where A is an event. The count branch is part of the main model logic rather than a purely isolated branch that outputs total PE count. Its output directly changes the behavior of the time decoder and therefore the final reconstructed PE time profile.

\subsection{Time branch: count-conditioned query decoding}

The time branch follows a query-decoder design. The decoder maintains a learnable query bank with maximum capacity \(Q_{\max}=20\). The key point is that the query-bank capacity is fixed, whereas the number of active queries varies from waveform to waveform. For each active query \(q\), the model predicts
\begin{itemize}
\item photon / no-photon classification logits;
\item a one-dimensional mask logit sequence over time bins;
\item a time estimate derived from the mask distribution.
\end{itemize}

If \(m_q(t)\) denotes the mask logits for query \(q\), we convert them into a normalized distribution using a softmax temperature \(\tau_m\),
\begin{equation}
p_q(t) = \frac{\exp(m_q(t)/\tau_m)}{\sum_{t'} \exp(m_q(t')/\tau_m)},
\end{equation}
and the corresponding time estimate is
\begin{equation}
\hat{\tau}_q = \sum_{t=1}^{L} p_q(t)\,\frac{t-1}{L-1}.
\end{equation}

The decoder is also count conditioned. The total PE number information is injected into the query embeddings, and the number of active queries is determined by the count branch rather than being fixed a priori. In the benchmark setup, the active query budget is given by
\begin{equation}
K_{\mathrm{act}} = \min\!\left(Q_{\max}, \hat{N}_{\mathrm{PE}} + \Delta\right),
\end{equation}
with \(\Delta=2\). In this way, \(Q_{\max}\) is the only hard constraint, while the effective number of queries used for reconstruction adapts to the predicted occupancy.

This mechanism matters mostly in high-pileup waveforms. When the predicted occupancy is small, the model naturally suppresses redundant queries, which reduces fragmentation in the reconstruction. When the occupancy is large, more queries are allowed to participate, increasing the capacity available for resolving pile-uped signals.

Let \(\alpha_q\) denote the photon probability of query \(q\), and \(\gamma\) the confidence threshold. If \(\mathcal{K}_{\mathrm{act}}\) denotes the active-query set, the accepted queries are
\begin{equation}
\mathcal{Q} = \{q \in \mathcal{K}_{\mathrm{act}} \mid \alpha_q \ge \gamma \},
\end{equation}
with \(\hat{m}=|\mathcal{K}|\) accepted hits. The final reconstructed PE-time profile \(\widehat{\npevst}\) is then obtained by assigning each accepted query to its most likely time bin,
\begin{equation}
\hat{y}_t = \sum_{q\in \mathcal{K}} \mathbb{I}\!\left[t = \arg\max_{t'} m_q(t')\right].
\end{equation}

This representation is used both for monitoring and as an auxiliary dense-profile supervision signal during training.

From a modeling perspective, the time branch is not purely a dense regressor, nor a fixed-size set predictor. It sits somewhere in between the two. The query decoder first produces a pool of candidate PE time hypotheses, and the count conditioning effectively controls how many of them are allowed to be active, depending on the predicted occupancy. The final PE-time profile is then formed by assembling the selected queries back into the reconstructed $\widehat{\npevst}$ representation.

This hybrid setup is essentially what allows \phast{} to stay consistent across different levels of pileup: it can still produce meaningful waveform-level count and time estimates, while also yielding the PE-time profile required by the task.

\subsection{Target construction, matching, and loss function}

For training, the waveform target is first converted into a set-based representation consisting of individual hit positions and corresponding local masks. Each ground-truth hit is assigned a Gaussian mask with \(\sigma=1.5\) ns and support radius of \(6\) ns, which provides a smooth supervision signal for both localization and shape learning.

Given this representation, we perform bipartite matching between predicted queries and target hits using a Hungarian algorithm. The matching cost combines classification, mask, and time terms,
\begin{equation}
\mathcal{C}_{ij}
=
\lambda_{\mathrm{cls}}\,\mathcal{C}^{\mathrm{cls}}_{ij}
+ \lambda_{\mathrm{mask}}\,\mathcal{C}^{\mathrm{BCE}}_{ij}
+ \lambda_{\mathrm{dice}}\,\mathcal{C}^{\mathrm{Dice}}_{ij}
+ \lambda_{t}\,\left|\hat{\tau}_{i}-\tau_{j}\right|,
\end{equation}
with weights \(1.0\), \(1.0\), \(1.0\), and \(500.0\), respectively. The relatively large time weight reflects the fact that accurate time alignment is the dominant supervision signal in the matching stage.

Once the optimal assignment is obtained, the training objective is computed over the matched pairs together with several auxiliary terms that operate at different levels of the reconstruction. The full loss in the benchmark configuration is
\begin{equation}
\begin{aligned}
\mathcal{L} =
&\ \alpha_{\mathrm{cls}}\,\mathcal{L}_{\mathrm{cls}}
+ \alpha_{\mathrm{mask}}\,\mathcal{L}_{\mathrm{mask}}
+ \alpha_{\mathrm{dice}}\,\mathcal{L}_{\mathrm{dice}} \\
&+ \alpha_{t}\,\mathcal{L}_{t}
+ \alpha_{\mathrm{total\mbox{-}e}}\,\mathcal{L}_{\mathrm{total\mbox{-}e}}
+ \alpha_{\mathrm{recon}}\,\mathcal{L}_{\mathrm{recon}} \\
&+ \alpha_{\mathrm{EMD}}\,\mathcal{L}_{\mathrm{EMD}}
+ \alpha_{\mathrm{count}}\,\mathcal{L}_{\mathrm{count}}.
\end{aligned}
\end{equation}
with weights 5.0, 1.0, 1.0, 100, 0.05, 0.25, 0.05 and 1.0/0.5.
Here, \(\mathcal{L}_{\mathrm{cls}}\) is a focal-style photon/no-photon classification loss, while \(\mathcal{L}_{\mathrm{mask}}\) and \(\mathcal{L}_{\mathrm{dice}}\) supervise the matched query masks. The time loss \(\mathcal{L}_{t}\) enforces alignment between predicted and matched hit times. The term \(\mathcal{L}_{\mathrm{total\mbox{-}e}}\) constrains consistency between the decoder-implied total PE and the label, while \(\mathcal{L}_{\mathrm{recon}}\) provides dense supervision on the reconstructed waveform using a smooth-\(L_1\) objective. In addition, \(\mathcal{L}_{\mathrm{EMD}}\) imposes a distribution-level constraint based on cumulative profiles, and \(\mathcal{L}_{\mathrm{count}}\) corresponds to an ordinal binary cross-entropy loss on the TotalPEHead.

In practice, the count-loss weight is typically set in the range \(0.5\text{--}1.0\), and auxiliary count supervision from intermediate decoder layers can be included when additional stability is needed.

\section{Training dataset preparation and optimization}
\label{sec:training}

\subsection{Matched-dataset training protocol}

Each dataset is trained independently on its own waveform/PE-time profile pairs, using the same model family and a common set of optimization hyperparameters. We therefore treat the study as direct supervised training on the matched dataset for each of the five FastSlow and one Uniform cases. This design keeps the interpretation of the results straightforward: each trained model is evaluated on the same waveform regime on which it was optimized.

\subsection{Internal three-phase schedule}

Although the data source remains fixed within each matched-dataset run, the model optimization itself proceeds through three internal phases. These phases should be understood as optimization phases rather than different datasets.

\paragraph{Phase 1.}
The decoder parameters are frozen and only the counting pathway is optimized. In this phase, the count branch is supervised with the ground-truth waveform-level PE count, while time losses are inactive. The purpose of this phase is to stabilize occupancy estimation before activating full time reconstruction.

\paragraph{Phase 2.}
The time decoder is activated, while the count branch continues to be supervised with the ground-truth total PE count. The time decoder now uses the true count both as count-conditioning input and as the query-budget limit. In other words, time reconstruction is trained with teacher-forced occupancy information while the model learns to align query outputs to the target PE-time profile.

\paragraph{Phase 3.}
The count branch is still trained against the ground-truth total PE count, but the time decoder no longer uses the true count as its control input. Instead, it uses the predicted count, and the active query budget becomes
\begin{equation}
K_{\mathrm{act}} = \min(Q_{\max}, \hat{N}_{\mathrm{PE}} + 2).
\end{equation}
This phase therefore exposes the time branch to the same type of occupancy uncertainty that appears at inference stage.

The automatic switch from phase 1 to phase 2 depends on validation count performance together with a minimum-epoch requirement. Since the runs discussed here use a fixed matched dataset throughout training, the relevant minimum-epoch setting is the one associated with the active data-stage configuration of that run. The dataset-specific phase-1 thresholds used in the benchmark are summarized in table~\ref{tab:phase_switch}.

\begin{table}[t]
\centering
\renewcommand{\arraystretch}{1.1}
\begin{tabular}{lccc}
\hline
Dataset & Min epochs (P1$\rightarrow$P2) & Acc thr. & Tol. acc thr. \\
\hline
FastSlow(20/100) & 15 & 0.70 & 0.95 \\
FastSlow(30/100) & 15 & 0.75 & 0.95 \\
FastSlow(40/100) & 20 & 0.80 & 0.95 \\
FastSlow(50/100) & 8  & 0.80 & 0.95 \\
FastSlow(80/200) & 8  & 0.80 & 0.95 \\
\hline
\end{tabular}
\caption{Dataset-specific thresholds for switching from phase 1 to phase 2 in matched-dataset training.}
\label{tab:phase_switch}
\end{table}

The automatic switch from phase 2 to phase 3 is shared across the FastSlow runs and is triggered when
\begin{equation}
\texttt{dt\_rms} < 2.0, \qquad
\texttt{match\_rate} \ge 0.95, \qquad
\texttt{acc\_mean} \ge 0.5.
\end{equation}
These conditions reflect the intended logic of the schedule: the model first learns occupancy, then learns time with teacher-forced occupancy, and only afterward is asked to reconstruct time under its own predicted count budget. Although some configuration files also store a profile-MAE threshold, the actual phase-2 to phase-3 switch in the present code path is governed by the three conditions above.

\subsection{Optimization setup}

The optimization setup uses AdamW with learning rate \(5\times 10^{-4}\), weight decay \(10^{-4}\), cosine annealing over \(100\) epochs, gradient clipping at \(1.0\), and batch size \(1024\). The run-level configuration summary is given in table~\ref{tab:config}. These settings are common across the matched-dataset benchmark discussed below.

\begin{table}[t]
\centering
\begin{tabular}{ll}
\hline
Setting & Value used in the reported runs \\
\hline
Dataset regime & matched-dataset training, one run per dataset \\
Waveform length & \(500\) bins/ns \\
Waveform/PE-time profile pairs per domain & \(100{,}000\) \\
Backbone dimension & \(128\) \\
Attention heads & \(4\) \\
Encoder layers & \(4\) \\
Decoder layers & \(2\) \\
Feed-forward dimension & \(256\) \\
Patch size / stride & \(5\) / \(1\) \\
Number of queries & \(20\) \\
Count classes & \(11\) \\
Batch size & \(1024\) \\
Validation ratio & \(0.2\) \\
Optimizer & AdamW \\
Initial learning rate & \(5\times 10^{-4}\) \\
Scheduler & CosineAnnealingLR \\
Evaluation size & about \(2\times 10^4\) waveform/PE-time profile pairs per dataset \\
\hline
\end{tabular}
\caption{Hyperparameter configuration for the benchmark.}
\label{tab:config}
\end{table}

\section{Performance}
\label{sec:eval}

\subsection{Metrics and evaluations}

We evaluate the method with count and time metrics, respectively. For each of the 11 count classes, we, on the one hand, choose classwise exact accuracy, and on the other hand, take a tolerated accuracy, which means tolerance for confusion between most adjacent classes. Count relative error is also taken into account and is one of the most important indicators.

For time metrics, let \(\mathcal{H}_i=\{\tau_{i,k}\}_{k=1}^{n_i}\) be the ground-truth normalized hit-time set and let \(\widehat{\mathcal{H}}_i=\{\hat{\tau}_{i,\ell}\}_{\ell=1}^{\hat{m}_i}\) be the accepted predicted hit-time set after confidence thresholding. We construct a Hungarian matching between \(\widehat{\mathcal{H}}_i\) and \(\mathcal{H}_i\), then retain only matched pairs satisfying
\begin{equation}
|\hat{\tau}_{i,\ell}-\tau_{i,k}| \le \tau_{\mathrm{match}},
\end{equation}
where \(\tau_{\mathrm{match}}\) is the normalized matching window. If \(\mathcal{M}_i\) denotes the set of retained matched pairs for waveform/PE-time profile pair \(i\), then the matched-hit rate is $\frac{\sum_i |\mathcal{M}_i|}{\sum_i  n_i}$. Using the ns unit, we calculate each matched time residual and its mean and root mean square error (RMS).

The containment metrics: the $x$ quantiles of  \(|\Delta t|\) distribution over all matched pairs are denoted like \texttt{dt\_contain\_x}.
In this work, time metrics are the primary figures of merit, though a qualified method needs to take into account both the performance of count and that of time.
Each dataset is evaluated with its matched model on about \(20{,}000\) waveform/PE-time profile pairs. Table~\ref{tab:results-main} summarizes the main metrics.

\begin{table}[t]
\centering
\resizebox{\textwidth}{!}{
\begin{tabular}{lccccccc}
\hline
Dataset & \texttt{acc\_mean} & \texttt{tolerated\_acc\_mean} & \texttt{count\_mae} & \texttt{match\_rate} & \texttt{dt\_rms} & \texttt{dt\_contain\_68} & \texttt{dt\_contain\_90} \\
\hline
FastSlow(20/100) & 0.768 & 0.990 & 0.244 & 0.997 & 0.839 & 0.778 & 1.397 \\
FastSlow(30/100) & 0.801 & 0.993 & 0.206 & 0.996 & 0.875 & 0.843 & 1.462 \\
FastSlow(40/100) & 0.819 & 0.994 & 0.188 & 0.995 & 0.876 & 0.837 & 1.465 \\
FastSlow(50/100) & 0.828 & 0.994 & 0.178 & 0.995 & 0.883 & 0.849 & 1.481 \\
FastSlow(80/200) & 0.866 & 0.997 & 0.135 & 0.995 & 0.886 & 0.833 & 1.496 \\
Uniform & 0.924 & 0.998 & 0.078 & 0.997 & 0.928 & 0.887 & 1.562 \\
\hline
\end{tabular}}
\caption{Main matched-dataset benchmark results for both FastSlow  and Uniform datasets. Tolerated count accuracy remains close to unity and time RMS stays below one bin throughout the benchmark.}
\label{tab:results-main}
\end{table}

\subsection{Count reconstruction performance}

The count-related metrics reveal a noticeable difference between exact count recovery and tolerant count recovery. Exact count accuracy is close to unity at low occupancy, but gradually decreases as total PE number increases, as shown in the left panel of figure~\ref{fig:benchmark_compare}. In comparison, the tolerated count accuracy in the middle panel remains relatively stable, suggesting that most prediction errors correspond to small local deviations rather than large failures.

This trend is also consistent with the matched-dataset summary table. Even for the more challenging FastSlow configurations, the mean of tolerated accuracy stays very close to one, whereas the mean of accuracy shows a stronger dependence on pileup level. In practice, this indicates that the count branch is still able to capture the overall occupancy reliably, although recovering the exact integer count becomes more difficult when many photons arrive or many PEs are generated within a narrow time interval.

\begin{figure}[t]
\centering
\includegraphics[width=0.48\textwidth]{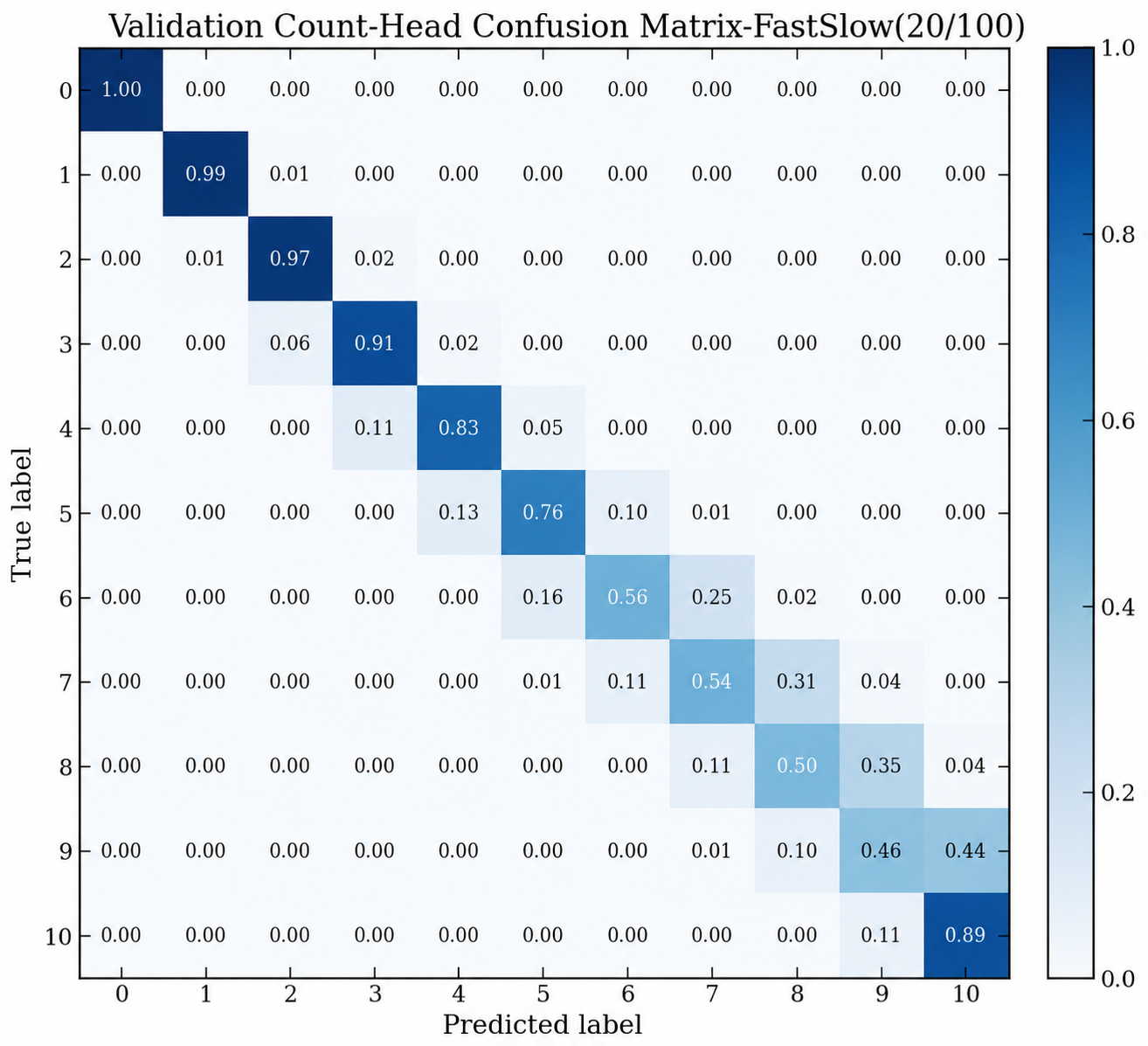}
\hfill
\includegraphics[width=0.48\textwidth]{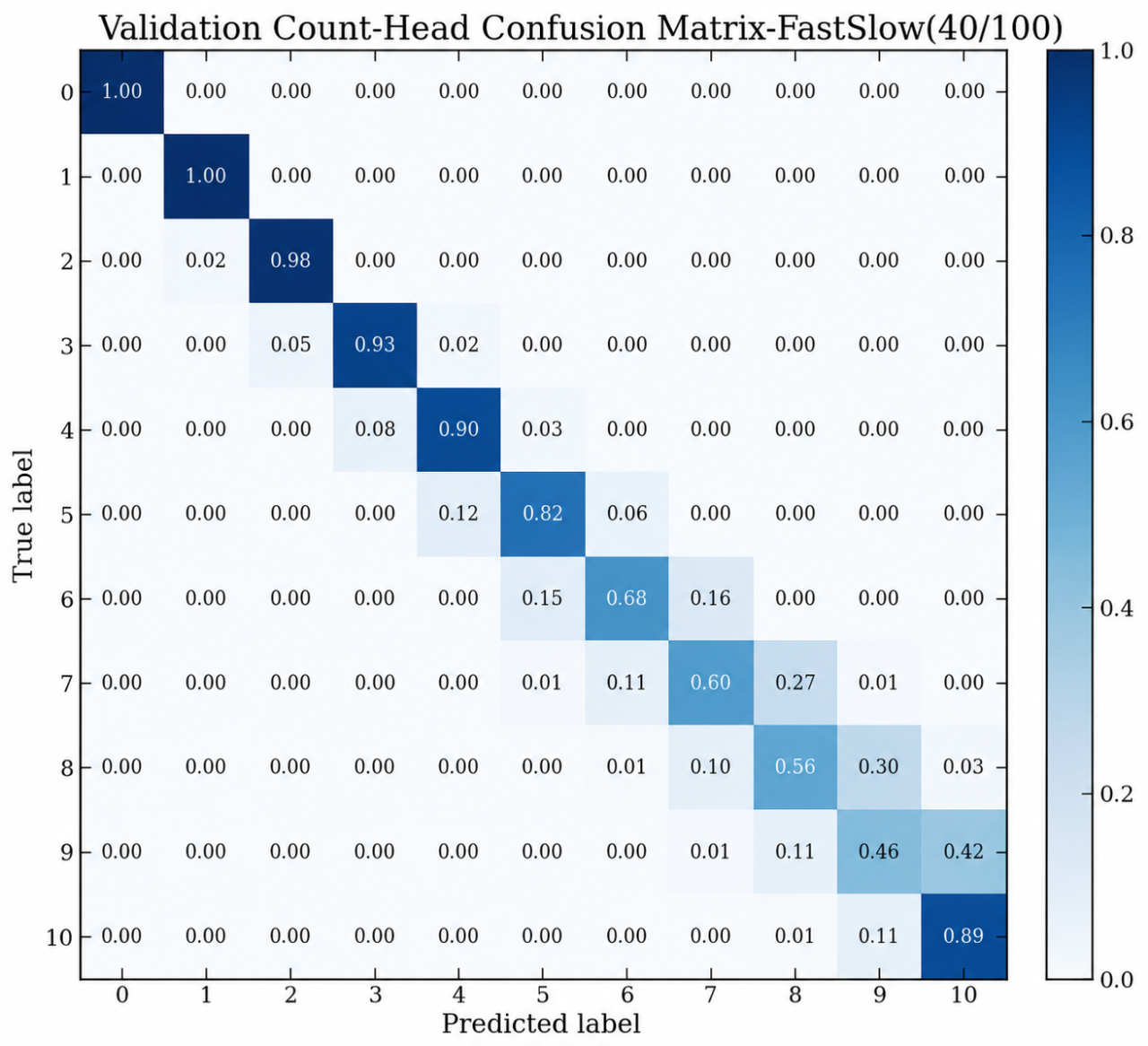}
\vspace{0.6em}
\includegraphics[width=0.48\textwidth]{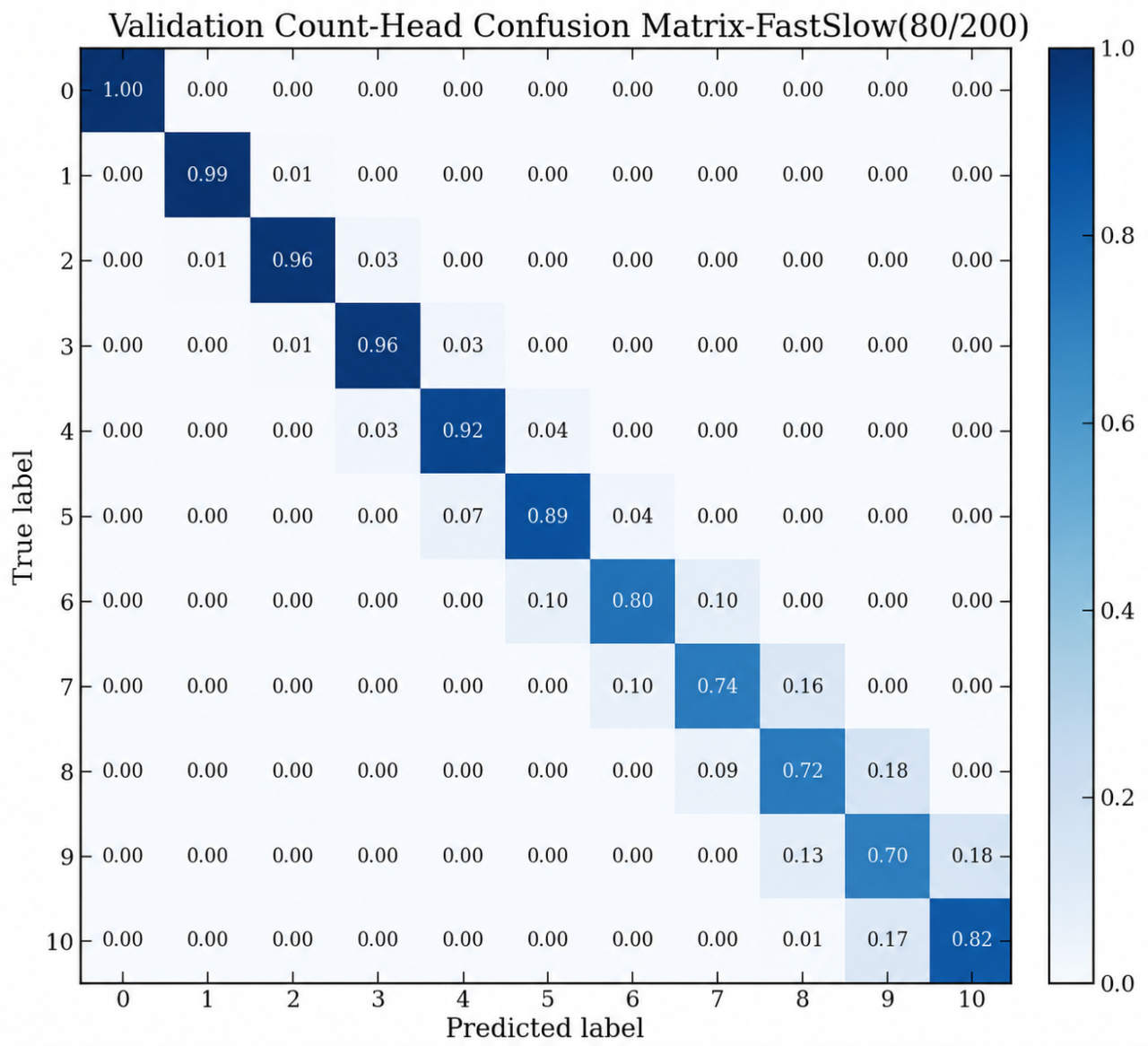}
\hfill
\includegraphics[width=0.48\textwidth]{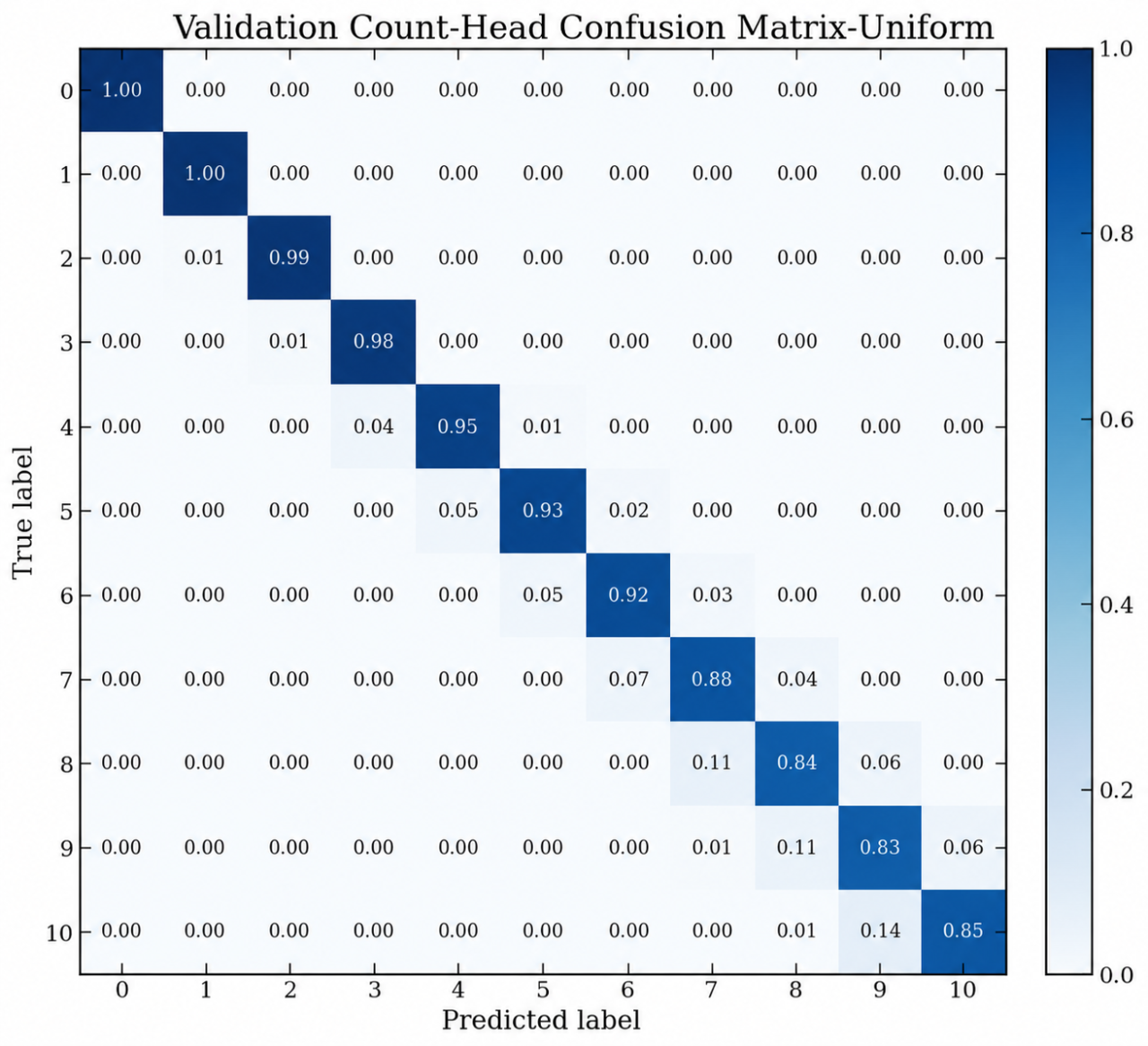}
\caption{TotalPEHead confusion matrices for three representative FastSlow datasets and Uniform dataset. \textbf{Top Left}: \textbf{Fastslow}(20/100); \textbf{Top Right}: \textbf{Fastslow}(40/100); \textbf{Bottom Left}: \textbf{Fastslow}(80/200); \textbf{Bottom Right}: \textbf{Uniform}. In all cases, the major samples remain concentrated near the diagonal, while the off-diagonal spread grows for severe pileup. }
\label{fig:count_confusion}
\end{figure}

A similar pattern can be seen in the count-head confusion matrices in figure~\ref{fig:count_confusion}. Most entries remain clustered near the diagonal, indicating that the predicted occupancy is usually close to the true value, with errors typically limited to only one or a few PE in high-pileup cases.

\subsection{Time reconstruction performance}

Time performance is considerably more stable than exact count recovery. Across all six datasets, the matched-hit rate stays consistently high, while the time bias, characterized by mean residual time, remains very small. In the benchmark results, mean residual time varies only from (0.0021) ns for FastSlow((20/100)) to (0.0163) ns for FastSlow((80/200)), indicating that the reconstruction introduces little systematic time shift overall.

The time spread is also reasonably well controlled. The RMS of residual time ranges from (0.839) to (0.928) ns, and the (68\%) containment remains between (0.778) and (0.887) ns throughout the benchmark. In other words, most matched hits are still reconstructed within roughly $1$ ns, even in the more difficult pileup conditions.

\begin{figure}[t]
\centering
\includegraphics[width=0.48\textwidth]{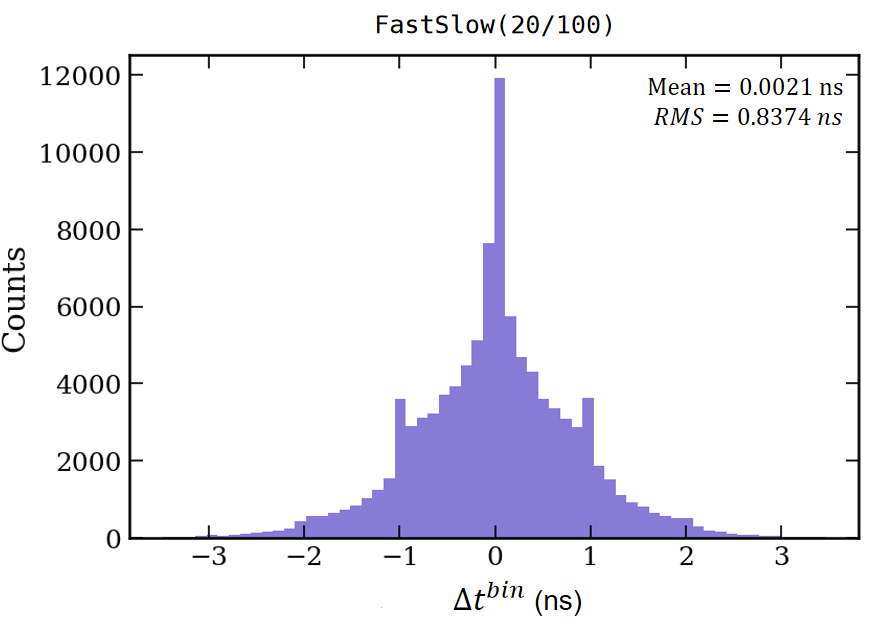}
\hfill
\includegraphics[width=0.48\textwidth]{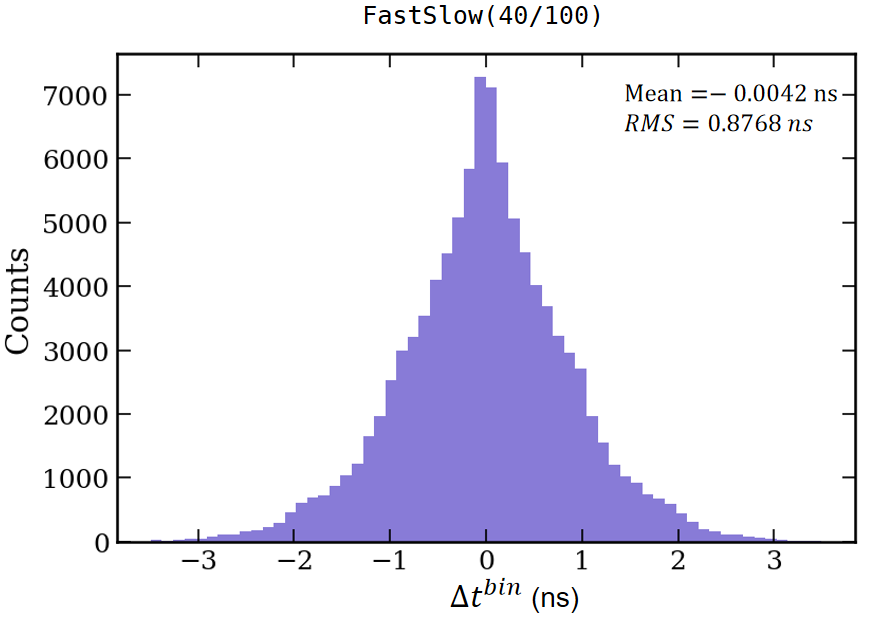}
\vspace{0.6em}
\includegraphics[width=0.48\textwidth]{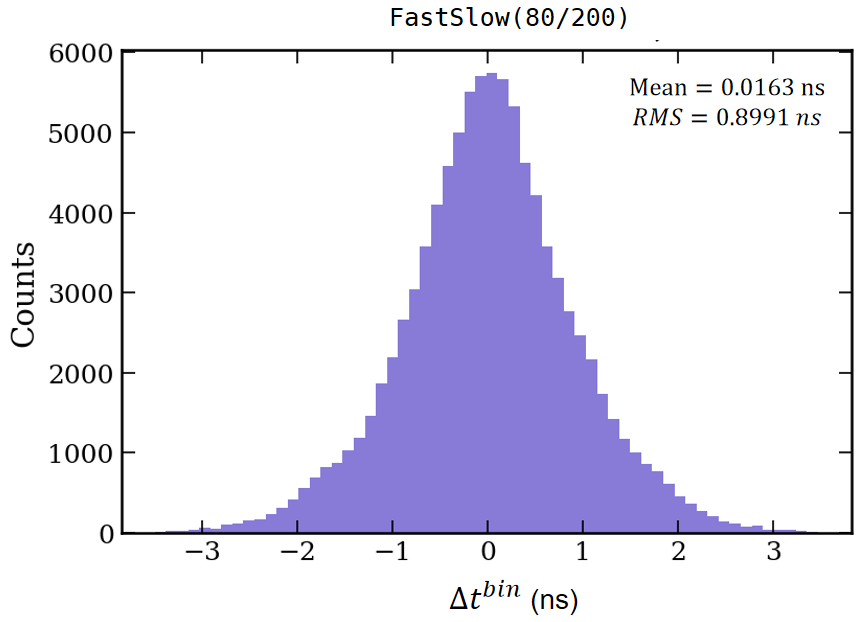}
\hfill
\includegraphics[width=0.48\textwidth]{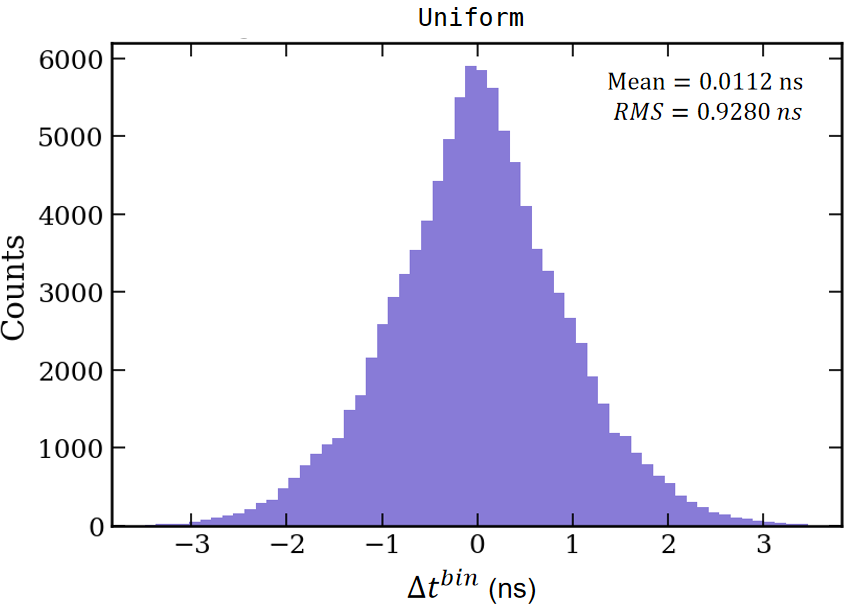}
\caption{Per-dataset residual time distribution in ns. 
\textbf{Top Left}: \textbf{FastSlow}(20/100) dataset; 
\textbf{Top Right}: \textbf{FastSlow}(40/100) dataset; 
\textbf{Bottom Left}: \textbf{FastSlow}(80/200) dataset; 
\textbf{Bottom Right}: \textbf{Uniform} dataset. 
The near-zero mean value indicates that the model does not have any systematic bias. 
The RMS value is approximately 0.8--0.9\,ns, which is close to that of TTS.}
\label{fig:dt_histograms}
\end{figure}

This trend can also be seen in the right panel of figure~\ref{fig:benchmark_compare}, where the time resolution worsens gradually as the pileup level increases. Even for the hardest datasets, however, the matched models remain below about $1$ ns RMS. The same behavior appears in the residual time histograms in figure~\ref{fig:dt_histograms}: the central peak becomes broader for harder datasets, but the distributions do not develop large asymmetric tails or other obviously unstable features. Overall, the results suggest that time localization remains relatively reliable, even when exact count reconstruction starts to degrade under strong pileup.

\subsection{Representative reconstruction examples}
\begin{figure}[t]
\centering
\includegraphics[width=0.48\textwidth]{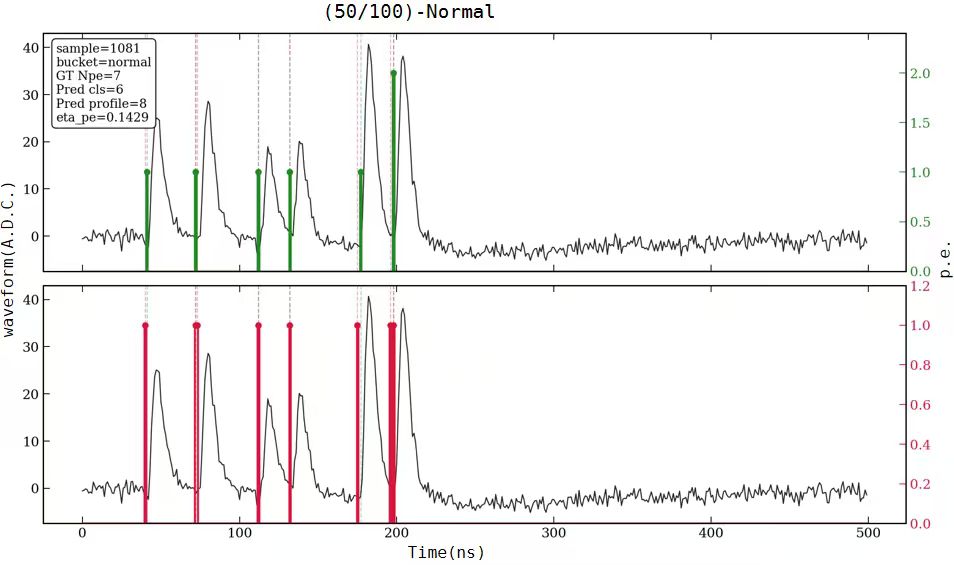}
\hfill
\includegraphics[width=0.48\textwidth]{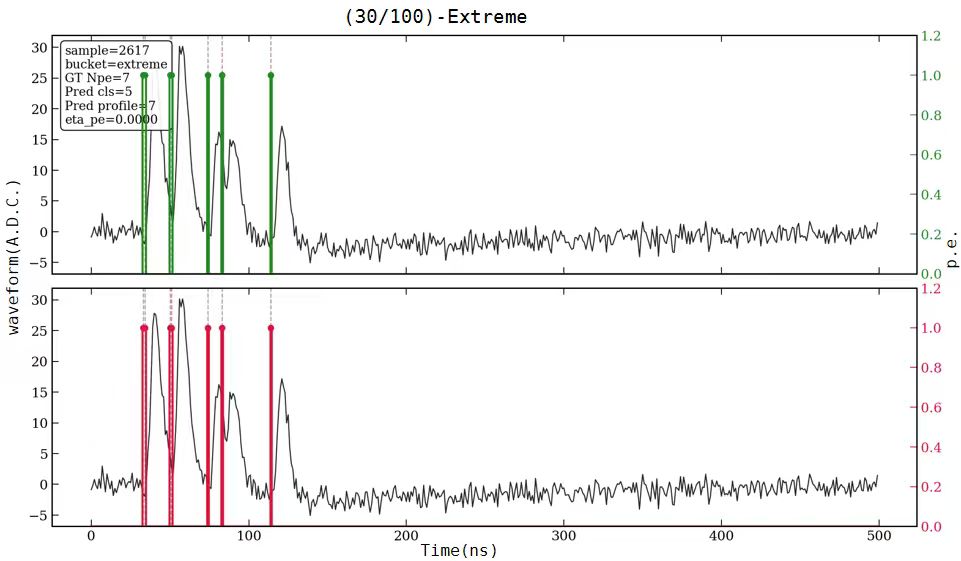}

\vspace{0.6em}

\includegraphics[width=0.48\textwidth]{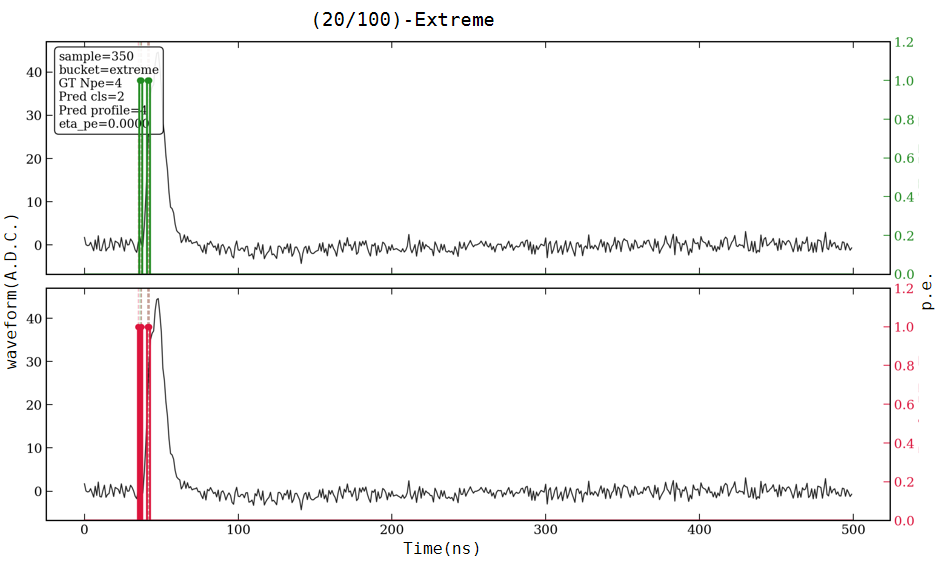}
\hfill
\includegraphics[width=0.48\textwidth]{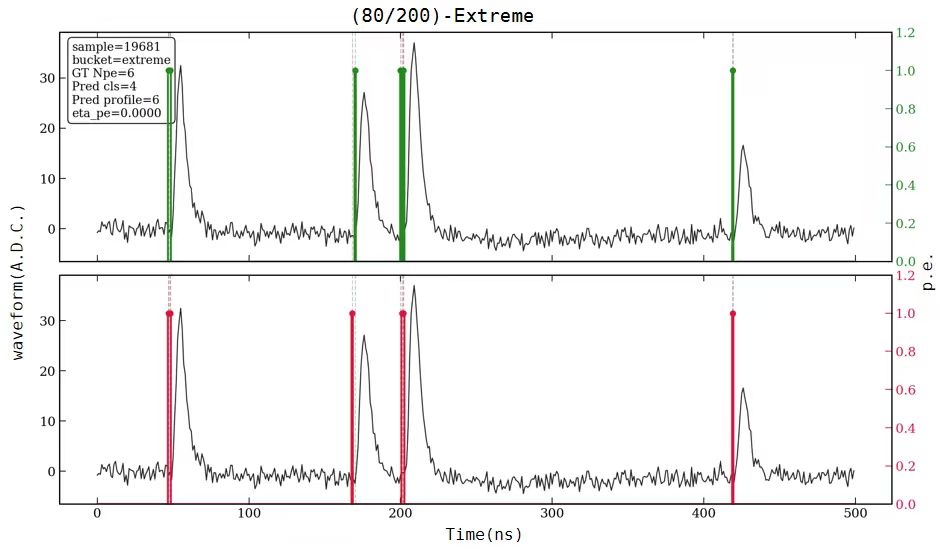}
\caption{Representative waveform reconstructions from the FastSlow benchmark. Here, eta\_pe in legend box denotes the relative error of the total PE count. 
\textbf{Top left}: a normal-bucket event from \textbf{FastSlow} (50/100); 
\textbf{Top right}: an extreme-bucket event from \textbf{FastSlow} (30/100); 
\textbf{Bottom left}: an extreme-bucket event from \textbf{FastSlow} (20/100); 
\textbf{Bottom right}: an extreme-bucket event from \textbf{FastSlow} (80/200). 
In each panel, the predicted profile follows the dominant time structures of the waveform, while the residual disagreement is mainly associated with local multiplicity ambiguity in severe pileup regions.}
\label{fig:example_reconstruction}
\end{figure}

Figure~\ref{fig:example_reconstruction} presents four representative reconstruction examples selected from different FastSlow datasets and difficulty buckets. Several common features can be observed across these cases.

First, the predicted time structures generally follow the main features of the waveform well, even when multiple PEs contributions overlap in time. Second, the remaining discrepancies are usually associated with local count mismatches rather than large time offsets. In many cases, the reconstruction may miss a nearby PE or merge two close contributions while still preserving the overall time pattern of the profile. Third, the harder examples are mainly characterized by stronger local pileup and poorer separation between neighboring peaks, which agrees with the dataset-dependent trends discussed earlier.

We do not focus on the visually best examples from the easy bucket, since many of those samples correspond to nearly empty or very low-occupancy waveforms and therefore provide limited information about the reconstruction behavior. The examples shown here are intended to highlight situations where the model must resolve temporal structures with pileup while still maintaining reasonable agreement with the target waveform.

\begin{figure}[t]
\centering
\includegraphics[width=0.32\textwidth]{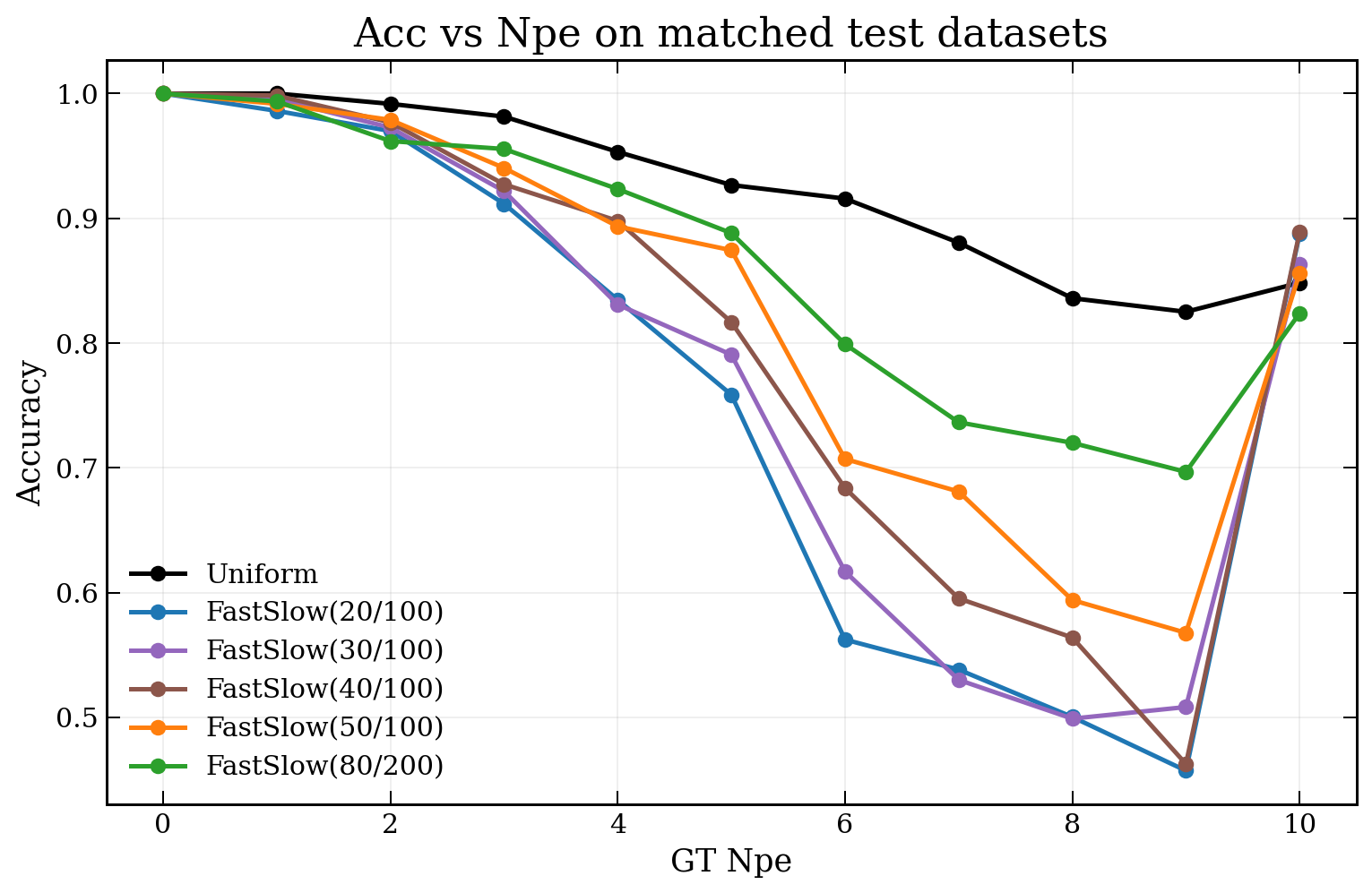}
\hfill
\includegraphics[width=0.32\textwidth]{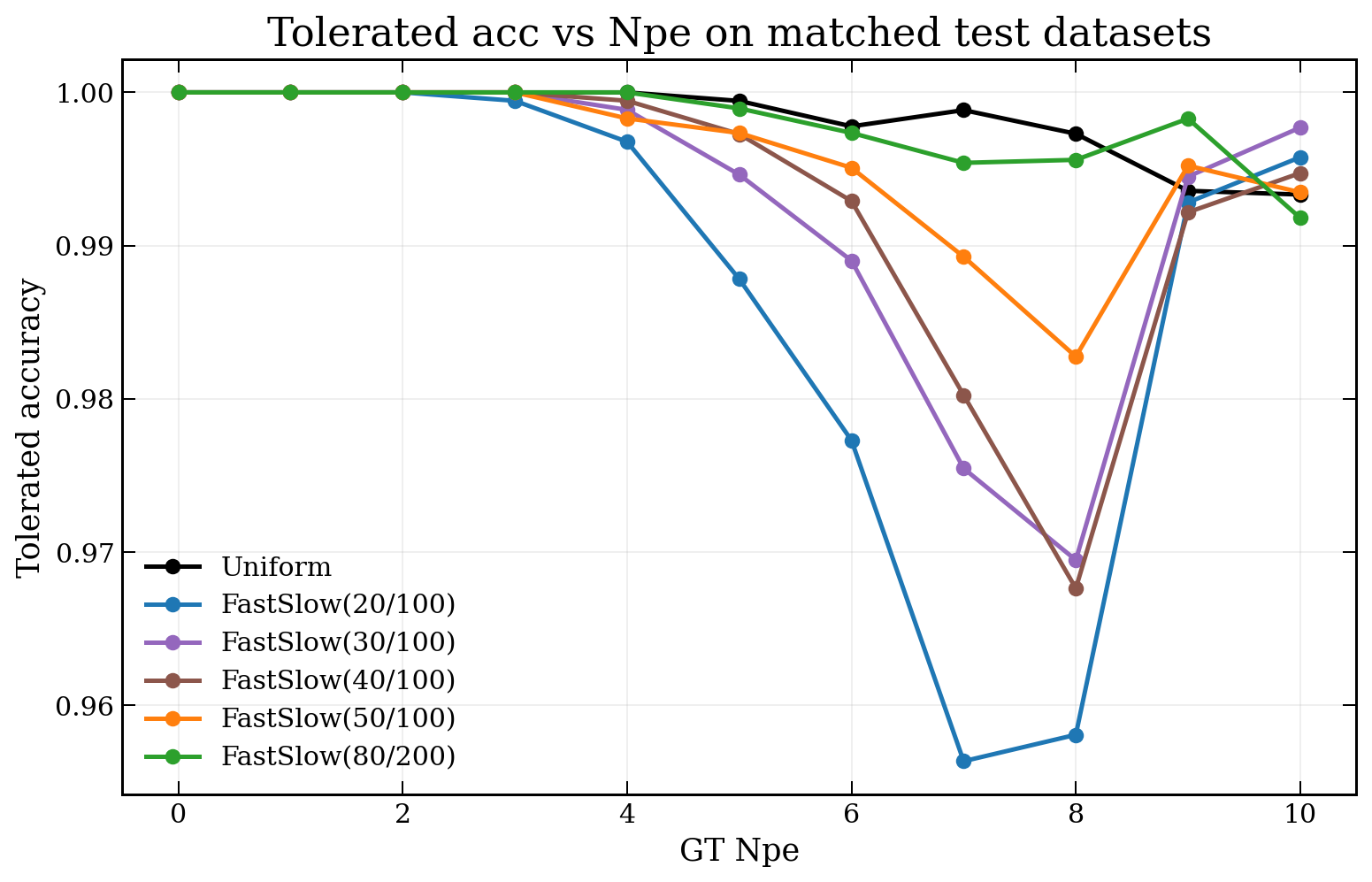}
\hfill
\includegraphics[width=0.32\textwidth]{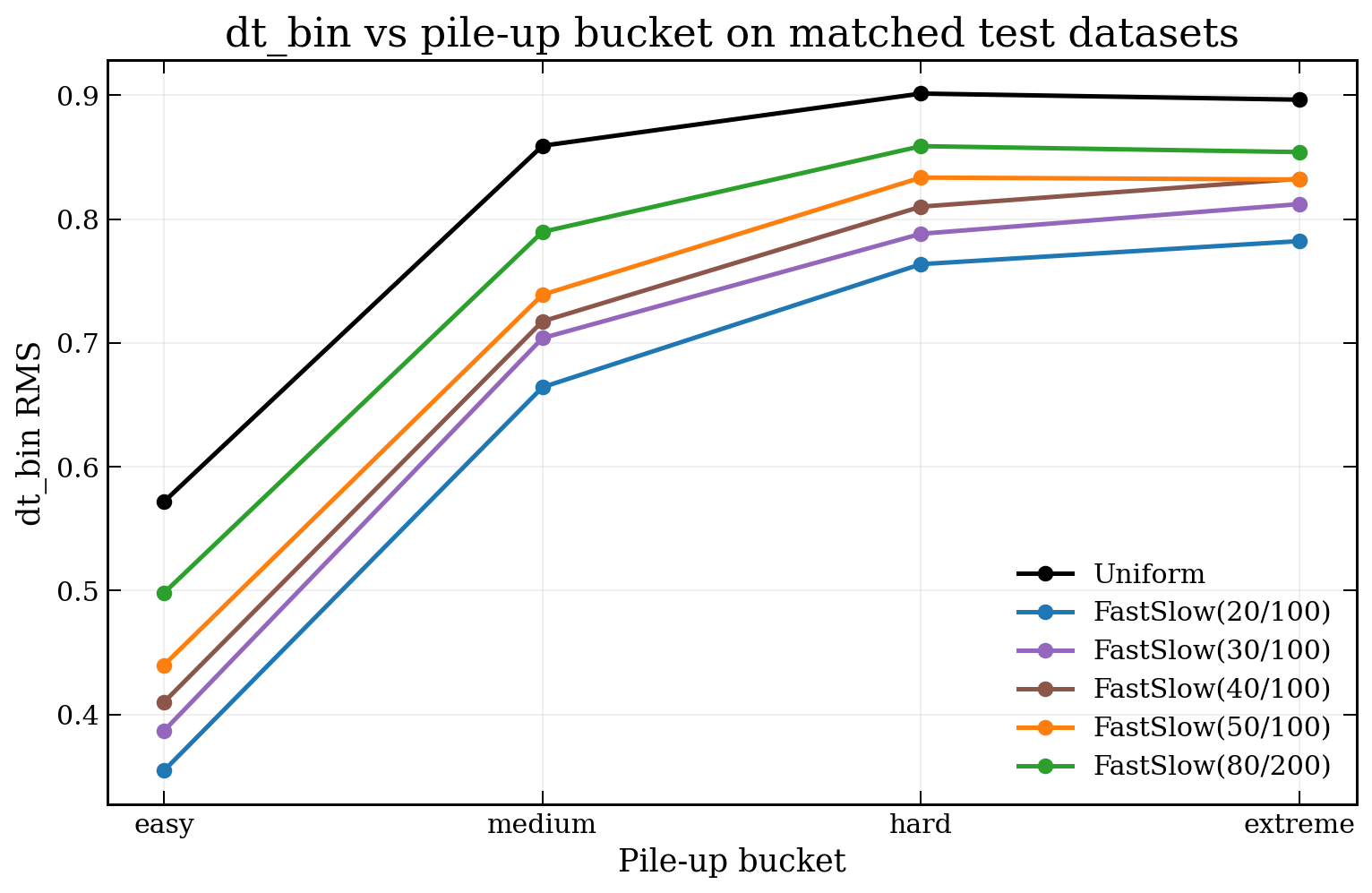}
\caption{Cross-dataset comparison plots within the FastSlow benchmark. \textbf{Left}: exact count accuracy as a function of ground-truth \(N_{\mathrm{PE}}\). \textbf{Middle}: tolerated count accuracy within a \(\pm 1\) PE window as a function of ground-truth \(N_{\mathrm{PE}}\). \textbf{Right}: time RMS as a function of the pileup bucket.}
\label{fig:benchmark_compare}
\end{figure}

Figure~\ref{fig:benchmark_compare} refines this picture. The left panel shows that exact count accuracy is close to unity at small occupancy and falls systematically at larger \(N_{\mathrm{PE}}\), with the main difficulty concentrated around \(N_{\mathrm{PE}}\gtrsim 6\). The middle panel shows that once a \(\pm 1\) PE tolerance is allowed, count performance remains substantially more stable across occupancy. The right panel shows that time quality degrades smoothly with pileup severity, but all matched FastSlow models remain below roughly $1$ ns RMS even in the hardest bucket shown.

\section{Discussion}
\label{sec:discussion}

The performance results lead to several observations. First, the separation between total PE number estimation and query-based time reconstruction is reflected directly in the reconstruction behavior: matched-hit rates remain consistently high across all five FastSlow datasets and one Uniform datasets, while the main performance loss appears in exact count recovery. Second, training on matched datasets alone is already enough to obtain stable performance over a range of FastSlow time configurations. Third, the FastSlow benchmark shows that time reconstruction remains relatively robust even in strong pileup conditions where recovering the exact PE multiplicity becomes more challenging. The qualitative examples show the same trend: the dominant time structures are usually reconstructed correctly, and the remaining errors are more often related to count ambiguity than to large time offsets.

The benchmark also helps identify the main direction for further study. At this stage, the central issue is less about whether \phast{} can reconstruct the waveform structure, and more about how its performance compares with conventional reconstruction methods and with simplified or ablated variants of the model. The current results also indicate that improving high-occupancy count recovery is likely to be more important than further optimizing time localization, since the time performance remains comparatively stable even when exact count accuracy begins to degrade.

\section{Conclusions}
\label{sec:conclusion}
In this paper, We present \phast{}, an end-to-end framework for reconstructing discrete PE-time profiles directly from PMT waveforms. The method combines a shared waveform encoder, an ordinal TotalPEHead used for counting PE, and a count-conditioned query decoder whose active query budget is adjusted according to the predicted occupancy. To test this setup under controlled conditions, we build a toy waveform pipeline that includes SPE response, charge smearing, overshoot, baseline fluctuations, and noise, and use it to construct a set of FastSlow double-temporal-component-mixture datasets together with a Uniform dataset.

Across the six datasets, \phast{} consistently achieves high matched-hit rates, stable tolerated-count accuracy, and sub-ns to near-ns time RMS. Overall, the model does a good job of capturing the main time structure of the time profile, even in pileup-heavy cases. The remaining issues are fairly clear as well: most errors come from exact multiplicity recovery in regions with strong local pileup, where hits tend to be merged or slightly split rather than completely missed.

Taken together, the results point to two natural next steps. One is a direct comparison with more conventional or experiment-specific reconstruction baselines. The other is to improve performance in high-occupancy regions, most likely by tightening the interaction between occupancy estimation and local hit separation. More generally, the benchmark suggests that end-to-end waveform-to-profile reconstruction is a practical direction for handling realistic PMT waveform task.

\acknowledgments

\bibliographystyle{JHEP}
\bibliography{biblio}
\appendix
\section{Waveform simulation}
\label{simulation detail}
The waveform simulation mainly consists of four procedures,including PE-count and arrival-time sampling, SPE template and charge smearing, overshoot, baseline, and noise. 
\paragraph{PE-count and arrival-time sampling.}
For each waveform/profile pair, we first sample the total PE number from a discrete uniform distribution.
\begin{equation}
N_{\mathrm{PE}} \sim \mathrm{Uniform}\{0,1,\ldots,2\mu_{\mathrm{PE}}\},
\end{equation}
where \(\mu_{\mathrm{PE}}\) is the mean total PE number of the dataset. In the benchmark datasets considered here, \(\mu_{\mathrm{PE}}=5\), so the total PE count is bounded by \(10\).

Conditioned on \(N_{\mathrm{PE}}\), the photon arrival times in the benchmark datasets are sampled from a two-component exponential mixture,
\begin{equation}
p(t) = r \frac{1}{\tau_f} e^{-t/\tau_f} + (1-r)\frac{1}{\tau_s} e^{-t/\tau_s},
\qquad r=0.7,
\end{equation}
with \(70\%\) fast component and \(30\%\) slow component. The five FastSlow datasets are defined by the parameter pairs
\((\tau_f,\tau_s)\) ns, respectively. In all cases, sampled times are shifted into the allowed signal window before filling the integer-valued target profile \(\npevst\).Besides,a Uniform  dataset is also introduced.

\paragraph{SPE template and charge smearing.}
Each PE is converted into a waveform contribution through an asymmetric SPE template. The active template is a Daya Bay style log-normal pulse shape~\cite{Jetter_2012},
\begin{equation}
s(t;t_0,A) =
\begin{cases}
A \exp\left[-\frac{1}{2}\left(\frac{\ln((t-t_0)/\tau_{\ln})}{\sigma_{\ln}}\right)^2\right], & t>t_0,\\
0, & t \le t_0,
\end{cases}
\end{equation}
where \(t_0\) is the PE start time after TTS jitter, \(\tau_{\ln}=7.0\), and \(\sigma_{\ln}=0.45\) in the benchmark configuration. The jitter itself is modeled as
\begin{equation}
\delta t_{\mathrm{TTS}} \sim \mathcal{N}(0,\sigma_{\mathrm{TTS}}^2),
\qquad \sigma_{\mathrm{TTS}} = 1.0~\mathrm{ns},
\end{equation}
so that the effective pulse start time is \(t_0+\delta t_{\mathrm{TTS}}\).

The SPE amplitude is proportional to the sampled single-PE charge. We model the charge smearing with a Gamma distribution.
\begin{equation}
q \sim \Gamma(k,\theta),
\qquad
k=\left(\frac{q_0}{\sigma_q}\right)^2,
\qquad
\theta=\frac{\sigma_q^2}{q_0},
\end{equation}
with \(q_0=1.0\) and \(\sigma_q=0.30\). The pulse amplitude is then
\begin{equation}
A = U_0 \, G \, q \, C_{\mathrm{ADC}},
\end{equation}
where \(U_0\) is the nominal SPE peak scale, \(G\) is the gain factor, and \(C_{\mathrm{ADC}}\) is the ADC conversion factor.

\paragraph{Overshoot, baseline, and noise.}
We add overshoot as an independent negative response term aligned with the SPE waveform. Its active form is a sum of a Fermi-gated exponential tail and an optional Gaussian contribution,
\begin{equation}
O(t) = u_0 f_{\mathrm{F}}(t)\,e^{-t/\tau_{\mathrm{os}}}
+ u_1 \exp\left[-\frac{(t-t_g)^2}{2\sigma_g^2}\right],
\end{equation}
where \(u_0\) and \(u_1\) are negative amplitudes proportional to the SPE peak height. In the benchmark configuration, the dominant overshoot scale is set by an overshoot ratio of \(5\%\) and an exponential decay constant \(\tau_{\mathrm{os}}=150\) ns.

At the waveform level, we add an waveform-by-waveform fluctuating baseline,
\begin{equation}
b_{\mathrm{evt}} = b_0 + \delta b,
\qquad
\delta b \sim \mathcal{N}(0,\sigma_b^2),
\end{equation}
and white electronic noise,
\begin{equation}
n_{\mathrm{white}}(t) \sim \mathcal{N}(0,\sigma_{\mathrm{white}}^2).
\end{equation}
where $\sigma_{white}=2.1 $ A.D.C. The generator also supports optional dark-noise pulses generated as random SPE superpositions, although the matched benchmark runs discussed here use the standard filtered waveform pipeline without relying on dark noise as a dominant effect. The final raw waveform is obtained by summing all PE contributions together with baseline and noise terms.

\section{Difficulty-aware split and sampling}
\label{difficulty-split}
The benchmark does not split waveform/profile pairs only by total PE count. Instead, we use a difficulty-aware label designed to reflect several aspects of pileup simultaneously. For each pair, we compute
\begin{itemize}
\item the total PE count \(N_{\mathrm{PE}}\);
\item the minimum spacing between adjacent non-zero time bins, denoted \(\Delta t_{\min}\);
\item the maximum PE occupancy in a single bin, denoted \(n_{\max}^{\mathrm{bin}}\);
\item the maximum PE count inside a short local window, denoted \(d_{\mathrm{local}}\).
\end{itemize}

The difficulty label is built in two steps. First, a base bucket is assigned from the total PE count:
\begin{equation}
d_0 =
\begin{cases}
0, & N_{\mathrm{PE}} < 3,\\
1, & 3 \le N_{\mathrm{PE}} \le 5,\\
2, & 6 \le N_{\mathrm{PE}} \le 8,\\
3, & N_{\mathrm{PE}} \ge 9.
\end{cases}
\end{equation}
Second, this base difficulty is adjusted according to explicit local pileup rules. We define an additive correction \(\delta d\) from three sources:
\begin{itemize}
\item \textbf{minimum hit spacing:} if \(\Delta t_{\min} \ge 20\), the difficulty is reduced by \(1\); if \(3 \le \Delta t_{\min} < 8\), it is increased by \(1\); if \(1 \le \Delta t_{\min} < 3\), it is increased by \(2\);
\item \textbf{same-bin pileup:} if \(n_{\max}^{\mathrm{bin}} \ge 2\), add \(1\); if \(n_{\max}^{\mathrm{bin}} \ge 3\), add another \(1\);
\item \textbf{local density:} if \(d_{\mathrm{local}} \ge 4\), add \(1\); if \(d_{\mathrm{local}} \ge 6\), add another \(1\).
\end{itemize}
Equivalently, one may write
\begin{equation}
\delta d =
\delta d_{\min \Delta t}
+ \mathbb{I}[n_{\max}^{\mathrm{bin}} \ge 2]
+ \mathbb{I}[n_{\max}^{\mathrm{bin}} \ge 3]
+ \mathbb{I}[d_{\mathrm{local}} \ge 4]
+ \mathbb{I}[d_{\mathrm{local}} \ge 6],
\end{equation}
with
\begin{equation}
\delta d_{\min \Delta t} =
\begin{cases}
-1, & \Delta t_{\min} \ge 20,\\
0, & 8 \le \Delta t_{\min} < 20 \ \text{or no valid pair exists},\\
1, & 3 \le \Delta t_{\min} < 8,\\
2, & 1 \le \Delta t_{\min} < 3.
\end{cases}
\end{equation}
The final label is clipped to four levels,
\begin{equation}
d = \mathrm{clip}(d_0 + \delta d, 0, 3),
\end{equation}
with the final bucket assignment defined by
\begin{equation}
d=0 \rightarrow \textit{easy}, \qquad
d=1 \rightarrow \textit{normal}, \qquad
d=2 \rightarrow \textit{hard}, \qquad
d=3 \rightarrow \textit{extreme}.
\end{equation}
In practice, this score is more informative than total PE count alone because it encodes both occupancy and local pileup.

We use this difficulty bucket in two places. First, the train/validation split is stratified by the bucket label so that all pileup regimes remain represented in both subsets. Second, the training dataloader uses balanced-batch sampling based on the same label, which reduces the tendency of optimization to be dominated by the easiest occupancy patterns. This is especially important in the FastSlow datasets, where waveforms with similar total PE number can still differ substantially in their local temporal structure.

\end{document}